\documentclass[aps,twocolumn,prd,showpacs,nofootinbib]{revtex4}
\usepackage{amsmath}
\usepackage{graphicx}
\usepackage{dcolumn}
\usepackage{bm}
\usepackage{amssymb}
\usepackage{latexsym}

\def\be{\begin{equation}}
\def\ee{\end{equation}}
\def\ba{\begin{eqnarray}}
\def\ea{\end{eqnarray}}

\begin{document}

\title{ On Primordial Perturbations of Test Scalar Fields
}

\author{Yun-Song Piao$^{a,c}$ and Yuan-Zhong Zhang$^{b,c}$}
\affiliation{$^a$College of Physical Sciences, Graduate School of
Chinese Academy of Sciences, YuQuan Road 19{\rm A}, Beijing
100049, China} \affiliation{$^b$CCAST (World Lab.), P.O. Box 8730,
Beijing 100080} \affiliation{$^c$Institute of Theoretical Physics,
Chinese Academy of Sciences, P.O. Box 2735, Beijing 100080, China}

\begin{abstract}

The primordial perturbations of test scalar fields not affecting
the evolution of background may be very interesting since they can
be transferred to the curvature perturbations by some mechanisms,
and thus under certain condition can be responsible for the
structure formation of observable universe. In this brief report
we study the primordial perturbations of test scalar fields in
various (super)accelerated expanding backgrounds.

\end{abstract}

\pacs{98.80.Cq} \maketitle

Due to the central role of primordial perturbations on the
formation of cosmological structure, it is very important to probe
their possible nature and origin. The primordial curvature
perturbations are generally supposed to originate from the vacuum
fluctuation of inflaton during the inflation \cite{BST}, see also
Ref. \cite{MC}. However, it is also possible that the curvature
perturbations are generated by the decay of curvaton field
\cite{LW, MT}, see also Ref. \cite{LM}, or the fluctuated coupling
constants during the reheating \cite{DGZK, K}, see also Ref.
\cite{T}, in which it is required that the perturbations of test
scalar fields not affecting the evolution of background during the
generation of perturbations can be transferred to the curvature
perturbations in the radiation after the reheating,
thus the final amplitude of the spectrum turns out not to be
directly related to inflation. This to some extent relaxes some
bounds on the inflation models and makes the building of models
more flexible \cite{DL}, see also Ref. \cite{P}.

The causal primordial perturbations may be generated in the
(super)accelerated expanding backgrounds, in which the
perturbations can leave the horizon during their generation and
then reenter the horizon after the transition which is followed by
standard FRW evolution. During the (super)accelerated expansion
$ah$ increases with time, where $h\equiv {\dot a}/a$ is the Hubble
parameter. Thus the evolution of scale factor can be simply taken
as $a(t) \sim t^n, (t\rightarrow \infty)$ with $n>1$ and $ a(t)
\sim (-t)^{n}, (t\rightarrow 0_-)$ with $n$ negative. Though
during the (super)accelerated expansion, the primordial
perturbations originated from the vacuum fluctuations of the
background field can be also produced, their spectra are generally
not scale invariant, see Ref. \cite{PZ2} for details. However, as
was mentioned, since the perturbations of test scalar fields can
be transferred to the curvature perturbations in the radiation by
some mechanisms, it can be very interesting to study the
primordial perturbations originated from the test scalar fields in
the (super)accelerated expanding backgrounds. The relevant studies
can also be seen in the earlier literatures, e.g. on PBB scenario
\cite{V}. In this brief report, we will briefly show our result,
and then discuss relevant issues and differences from those in the
previous studies.

We will assume that $n$ is a constant for simplicity, and work
with the parameter $\epsilon \equiv-{\dot h}/h^2$, which describes
the change of $h$ in unit of Hubble time and depicts the abrupt
degree of background change. Thus for above background we have
$\epsilon= 1/n < 1$. In the conformal time $\eta$, we obtain $
a(\eta) \sim (-\eta)^{1/(\epsilon -1)}$. Thus we have the scale
factor \be a\sim \left({1\over (1-\epsilon) a h}\right)^{1\over
\epsilon -1}.\label{ah}\ee
Note that the later the perturbation reenters the horizon at late
time, the earlier its generation will be.
Generally it may be convenient for our purpose to define \be {\cal
N} \equiv \ln\left({k_e\over k}\right)\equiv \ln\left({a_e h_e
\over a h}\right), \label{caln}\ee which measures the efolding
number of mode with some scale $\sim k^{-1}$ which leaves the
horizon before the end of the (super)accelerated expanding phase,
where the subscript `$e$' denotes the end time of the
(super)accelerated expanding phase, and thus $k_e$ is the last
mode to be generated, see Ref. \cite{KST}. When taking
$ah=a_0h_0$, where the subscript `0' denotes the present time, we
generally have ${\cal N}\sim 50$, which is required by observable
cosmology. From Eq.(\ref{ah}), we can obtain \be {a_e\over
a}=\left({a h\over a_e h_e}\right)^{1\over \epsilon -1}=
\left({k\over k_e}\right)^{1\over \epsilon -1}, \label{asim}\ee
which, combined with Eq.(\ref{caln}), induces $a_e/a= e^{{\cal
N}/( 1-\epsilon)}$. Thus we can see that the change of scale
factor is quite dependent on $\epsilon$, e.g. when $\epsilon\simeq
0$, the scale factor exponentially expands, which is the case of
usual inflation models, while for enough negative $\epsilon$, the
change of scale factor is generally very small, and further in the
limit $\epsilon\rightarrow -\infty$, the scale factor will be
nearly unchanged, which corresponds to the case in Ref.
\cite{Piao, Piaoii}. From Eqs.(\ref{caln}) and (\ref{asim}), we
can obtain \be {h_e\over h} =\left({a_e h_e\over a
h}\right)^{\epsilon\over \epsilon -1}= \left({k_e\over
k}\right)^{\epsilon\over \epsilon-1}. \label{hhe}\ee Thus we can
see that when $\epsilon\simeq 0$, $h$ is nearly unchanged, while
when $\epsilon\rightarrow -\infty$, we have $h_e/h\simeq k_e/k$,
whose logarithm is the efolding number. The Eq. (\ref{hhe})
will be used in the following.

For the test scalar field $\varphi$, in the momentum space, its
motion equation is given by \be u_k^{\prime\prime} +(k^2-f(\eta))
u_k = 0 ,\label{uki}\ee where $u_k$ is related to the perturbation
of $\varphi$ by $u_k \equiv a \varphi_k$ and the prime denotes the
derivative with respect to $\eta$, and $f(\eta)$ can be generally
written as $(v^2-1/ 4)/\eta^2$, in which $v$ is determined by the
evolution of background and the details of $\varphi$ field, such
as its mass, its coupling to the background.
When $v$ is approximately a constant, this equation is a Bessel
equation, whose general solutions are the Hankel functions.
In the regime $k\eta \rightarrow \infty $, all interesting modes
are very deep in the horizon of the (super)accelerated expanding
background, thus Eq.(\ref{uki}) can be reduced to the equation of
a simple harmonic oscillator, in which $u_k \sim e^{-ik\eta}
/(2k)^{1/2}$, which in some sense suggests that the initial
condition can be taken as usual Minkowski vacuum.
In the superhorizon scale, in which the modes become unstable and
grow, the expansion of Hankel functions to the leading term of $k$
gives \be u_k\simeq
{1\over \sqrt{2k}}(-k\eta)^{{1\over 2}-v} ,\label{uk}\ee where the
phase factor and the constant with order one have been neglected.

Unless $\epsilon \simeq 0$, during the (super)accelerated
expansion the Hubble parameter $h$ of the background will be
changed all along. Thus based on $1/\eta = (1-\epsilon) ah$ and
Eq.(\ref{uk}), the amplitude of perturbations of $\varphi$ can be
expected to continue to change after the corresponding
perturbations leaving the horizon, up to the end of the
(super)accelerated expanding phase. This suggests that in
principle we should take the value of $u_k$ at the time when the
(super)accelerated expanding phase ends to calculate the amplitude
of perturbations. The power spectrum of perturbations of scalar
field $\varphi$ is given by $
k^{3/2}|\varphi_k |\sim k^{3/2} |u_k(\eta_e)/ a_e | $. Thus from
Eq.(\ref{uk}),
we have \be
k^{3/2}|\varphi_k | \simeq  g\cdot \left({h_e\over 2\pi}\right)
\cdot \left({k\over k_e}\right)^{3/2-v}
,\label{pt}\ee where  $k_e=a_eh_e$ has been used and
$g=(1-\epsilon)^{v-1/2}$, which is obtained by using $1/\eta_e=
(1-\epsilon) a_eh_e$. From Eq. (\ref{pt}), it can be noticed that
generally different from the curvature perturbation on spatial
slice of uniform energy density, in which we can reasonably take
the value $h$ of Hubble parameter at the time when a given
wavelength $\sim k^{-1}$ crosses the horizon to calculate the
amplitude of perturbation since the comoving curvature
perturbation is nearly constant in the superhorizon scale, here we
have to take the value $h_e$ to do so, since the perturbations of
test scalar field is generally not expected to be constant but
change in the superhorizon scale up to the end of
(super)accelerated expanding phase. Thus when the spectrum
$k^{3/2}| \varphi_k | $ is scale invariant, i.e. $v \simeq 3/2$,
we obtain the amplitude $ \simeq g\cdot h_e/2\pi $, not $h/2\pi$
as usual. Further, it may be interesting to see that there is a
factor $g$ in Eq.(\ref{pt}), which can be neglected for $|\epsilon
|<1$ but can lead to an important magnifying of the amplitude of
perturbations when $\epsilon \rightarrow -\infty$. This was not
noticed in the earlier references.

Let us firstly check in which cases Eq. (\ref{pt}) can be
recovered to the familiar case, in which the value $h$ of Hubble
parameter at the time when a given wavelength crosses the horizon
may be used to calculate the amplitude of corresponding
perturbation. When the scalar field is massless, we have
$f(\eta)\equiv a^{\prime\prime}/a$, and thus
$v_{l}
=3/2 +{\epsilon\over (1-\epsilon)}$,
in which the subscript `$l$' denotes the quantity corresponding to
the massless scalar field. Combining it and Eq.(\ref{hhe}) into
Eq.(\ref{pt}), we have \be
k^{3/2}| \varphi_k |\simeq g\cdot \left({h_e\over 2\pi}\right)
\cdot \left({k\over k_e}\right)^{\epsilon\over \epsilon -1}=g\cdot
\left({h\over 2\pi}\right) ,\label{vkl}\ee  which is just the
usual result
used to the massless scalar field. 
Though there is a factor $g$ in Eq.(\ref{vkl}), it dose not affect
the amplitude significantly since in the massless case
$g=(1-\epsilon)^{v_l-1/2}=(1-\epsilon)^{1\over 1-\epsilon} \sim
{\cal O}(1)$ for arbitrary value of $\epsilon <1$. For the case of
inflation, in which $\epsilon\simeq 0$, which leads to $v\simeq
3/2 +\epsilon$ and $g=1$, we have the familiar result \be
k^{3/2}| \varphi_k |\simeq \left({h_e\over 2\pi}\right) \cdot
\left({k_e\over k}\right)^{\epsilon}={h\over 2\pi}. \ee This has
been discussed in Ref. \cite{LR} in the slow-roll approximation.

However, the case can be very different
for e.g. the massive scalar field. In this case,
we have \ba f(\eta)={a^{\prime\prime}\over a}-m_\varphi^2a^2 & = &
{v_l^2-1/4\over \eta^2}-{m_{\varphi}^2\over
h^2(1-\epsilon)^2}\cdot {1\over \eta^2}\nonumber\\ & = &
{v_m^2-1/4\over \eta^2},\label{fm}\ea where $ah
(1-\epsilon)=1/\eta$ has been used, and $v_m^2 =v_l^2 -
(m_{\varphi}/h(1-\epsilon))^2$,
the subscript `$m$' denotes the quantity corresponding to the
massive scalar field. Note that $h\sim 1/(-t)$ generally changes
with the time in the (super)accelerated background, which makes
$v_m$ not constant, thus Eq.(\ref{uki}) will be not the exact
Bessel equation, which makes us very difficult to obtain its
analytic solution. However, there are also some exceptions. If
there exists a non-minimally coupling $\sim R\varphi^2$ between
$\varphi$ and gravity, where $R \sim h^2$ is the Ricci curvature
scalar, we will have that $m^2_\varphi \sim R\sim h^2 $, and thus
$m^2_\varphi /h^2$ (and then $v_m$) is a constant.
Note that if $m_{\varphi}$ is far larger than $h(1-\epsilon)$,
$v_m$ will become imaginary, and thus can not be attached to the
spectrum index of $\varphi_k$, since in order to obtain the power
spectrum of $\varphi$ we need to take the module square of $u_k$
in Eq.(\ref{uk}). Thus the value of $m^2_{\varphi}$ should be
small positive, or negative for our purpose. The solution of
Eq.(\ref{uki}) with $f(\eta)$ given by Eq.(\ref{fm}), which
corresponds to replace $v$ in Eq.(\ref{pt}) with $v_m$, is \ba
k^{3/2} | \varphi_k | &\simeq & g\cdot \left({h_e\over
2\pi}\right) \cdot \left({k\over k_e}\right)^{3/2 -v_m} \nonumber
\\ &= & g\cdot \left({h\over 2\pi}\right) \cdot \left({k\over
k_e}\right)^{v_l -v_m} . \label{mpt}\ea where in the second line
Eq.(\ref{hhe}) has been used. Firstly, since generally $v_l\neq
v_m$
it seems that we can not take the value of $h$ at the time when a
given wavelength crosses the horizon to calculate the amplitude of
perturbations,
unless the exponent part of $k/k_e$ in Eq.(\ref{mpt}) can be
approximately cancelled, i.e. $v_l\simeq v_m$, which suggests
$m^2_{\varphi}/(1-\epsilon)^2 h^2 \ll 1$ and corresponds to the
massless case discussed in last paragraph.
The above results can be also understood as follows. For the
massless scalar field, in the superhorizon scale, i.e.
$k\eta\rightarrow 0$, we have
$u_k^{\prime\prime}-(a^{\prime\prime}/a)u_k \simeq 0$. Thus
$u_k\sim a$ is the solution of this equation. This means that in
the superhorizon scale $|\varphi_k |=|u_k/a| $ is a constant, thus
we can use the value $h$ of Hubble parameter at the time when a
given wavelength crosses the horizon to calculate the amplitude of
perturbation. However, for the massive scalar field, when
$k\eta\rightarrow 0$, we have $
u_k^{\prime\prime}-(a^{\prime\prime}/a-m_\varphi^2a^2)u_k \simeq
0$. This equation has one growing solution and one decay solution.
The growing solution is given by $ u_k\sim (-\eta)^{{1\over 2}-
v_m}\sim a^{(\epsilon-1)({1\over 2}- v_m)}$, where $ a(\eta) \sim
(-\eta)^{1/(\epsilon -1)}$ has been used. Thus generally one can
not obtain that the $|\varphi_k |=|u_k/a|$ is constant, which
ocurrs only when $(\epsilon-1)({1\over 2}- v_m)=1$. Taking $v_m$
defined in Eq.(\ref{fm}), one can find that $(\epsilon-1)({1\over
2}- v_m)=1$ corresponds to $v_m= v_l$, i.e. the massless case
$m_\varphi = 0$ .
We can take arbitrary values of $\epsilon<1$ to validate it. For
example taking $\epsilon= -1$ and $m ^2_{\varphi}/(1-\epsilon)^2
h^2 =-5/4 \neq 0$, we have $v_m=3/2$. Thus we can obtain
$u_k^{\prime\prime}-(2/\eta^2)u_k\simeq 0$, whose solution is
$u_k\sim \eta^{-1}$. Since $a\sim (-\eta)^{1/(\epsilon-1)}\sim
\eta^{-1/2}$, one can obtain $|\varphi_k |=|u_k/a |\sim a$, which
increases with the time. This means that $\varphi_k$ is generally
changed in the superhorizon scale. In principle this change will
last all along to the end of the (super)accelerated expanding
phase. Thus with this consideration it seems that in general case
we have to take the value at the time when the (super)accelerated
expanding phase ends to calculate the amplitude of perturbations.

Secondly, 
when $\epsilon\rightarrow -\infty$, we have
$g=(1-\epsilon)^{v_m-1/2} \simeq |\epsilon|^{v_m-1/2}$, and thus
\be k^{3/2}|\varphi_k|\simeq |\epsilon|^{v_m-1/2}\cdot {h_e\over
2\pi}\cdot \left({k\over k_e}\right)^{3/2-v_m}, \label{ge}\ee
From Eq.(\ref{ge}), we can see that if $v_m>1/2$ but $\not\simeq
1/2$, the amplitude of perturbations will be strongly magnified by
$\epsilon$ in the factor $g$. For example, for the scale invariant
spectrum in which $v_m=3/2$, we have the amplitude $\simeq
|\epsilon| h_e/2\pi$, in this case the magnification is
proportional to $\epsilon$. The above result is not difficult to
be understood. Note that as was mentioned in the definition, what
$\epsilon$ depicts is the abrupt degree of background change.
Generally the change of background will lead to the increase of
inhomogeneity, the more abrupt the change is, the more
inhomogeneous the background after the change is. $|\epsilon|\gg
1$ corresponds to a very abrupt change of background, thus it can
be expected that the amplitude of perturbations will obtain a
significant magnification related with the abrupt change of
background. This result means that for the (super)accelerated
expanding phase with $|\epsilon |\gg 1$, the $h_e$ (and thus the
energy scale) at the end time should be far lower than the case
with $\epsilon \simeq 0$ in order to make the amplitude of
perturbations responsible for the structure formation of
observable universe.

There have been some studies on the inflation with $\epsilon <0$,
e.g. the phantom inflation \cite{PZ, PZ1, GJ, BFM}, the ghost inflation \cite{ACMZ}. 
The extending to $\epsilon\rightarrow -\infty$ can be straight by
adjusting the parameters. In addition, in the island universe
model \cite{Piao, Piaoii}, in which initially the universe is in a
cosmological constant sea, then the local and abrupt background
changes (quantum fluctuations) violating the null energy condition
create some islands with matter and radiation, which under certain
conditions might corresponds to our observable universe, one also
take $\epsilon\rightarrow -\infty$ to describe the very abrupt
change of background. In this case $v_l^2\simeq 1/4$, and thus $m
^2_{\varphi}/(1-\epsilon)^2 h^2\simeq -2$ is required to make the
spectrum scale invariant, i.e. make $v_m=3/2$, the resulting
amplitude of perturbations is $|\epsilon|(h_e/2\pi)$, see also
Ref. \cite{DV} for a different study.


In summary, we studied the perturbations of test scalar fields not
affecting the evolution of background in the (super)accelerated
expanding backgrounds. In general case, the perturbations are
not unchanged in the superhorizon scale, 
thus in this note we use the value of $h$ at the time when the
(super)accelerated expanding phase ends to calculate the amplitude
of perturbations, as was showed in Eq.(\ref{pt}). More
interestingly, we note that there is a factor $g$ in
Eq.(\ref{pt}), which is relevant with the abrupt degree of
background change and may lead to a significant magnifying of the
amplitude of perturbations when the background changes very
abruptly, i.e. $|\epsilon|\gg 1$. Its origin are discussed in
detail here.
When considering the (nearly) massless scalar fields, the usual
case in which the amplitude of perturbation can be calculated by
taking the value $h$ of Hubble parameter at the time when a given
wavelength crosses the horizon can be recovered. In this case the factor
$g\sim {\cal O}(1)$ is negligible.  
This study may be interesting for the building of early universe
models dependent of test scalar field providing the primordial
perturbations.

\textbf{Acknowledgments} This work is supported in part by NNSFC
under Grant Nos: 10405029 and 90403032, as well as in part by
National Basic Research Program of China under Grant No:
2003CB716300, and in part by the Scientific Research Fund of
GUCAS(NO.055101BM03).

\end{document}